\documentclass[aps,preprint,amssymb,12pt,floatfix]{revtex4}
\usepackage{graphicx}
\usepackage{subfigure}
\usepackage{rotating} 
\usepackage{color}
\usepackage{amsmath,amsthm}
\usepackage{epstopdf}
\begin{document}
\title{On the importance of hydrodynamic interactions in the stepping kinetics of kinesin}
\author{Yonathan Goldtzvik$^1$, Zhechun Zhang$^1$, and D. Thirumalai$^{1,2}$}
%\maketitle
\affiliation{$^1$Biophysics Program, Institute for Physical Science and Technology and $^2$Department of Chemistry and Biochemistry, University of Maryland, College Park 20742, USA}
%\maketitle

\begin{abstract}
Conventional kinesin walks by a hand-over-hand mechanism on the microtubule (MT) by taking $\sim$ 8$nm$ discrete steps, and consumes one ATP molecule per step. The time needed to complete a single step is on the order of twenty microseconds. We show, using simulations of a coarse-grained model of the complex containing the two motor heads, the MT, and the coiled coil that in order to obtain quantitative agreement with experiments for  the stepping kinetics hydrodynamic interactions (HI) have to be included. In simulations without hydrodynamic interactions spanning nearly twenty $\mu$s  not a single step was completed in hundred trajectories. In sharp contrast, nearly 14\% of the steps reached the target binding site within 6$\mu$s when HI were included. Somewhat surprisingly, there are qualitative differences in the diffusion pathways in simulations with and without HI. The extent of movement of the trailing head of kinesin on the MT during the diffusion stage of stepping is considerably greater in simulations with HI than in those without HI. Our results suggest that inclusion of HI is crucial in the accurate description of motility of other motors as well.
\end{abstract}
%\date{\today}
\maketitle
\clearpage

\section{Introduction}
Conventional kinesin, to be referred to as kinesin from here on, is a molecular motor that plays a central role in cargo transportation in eukaryotic cells. In particular, it is involved in the transportation of vesicles and organelles as well as protein complexes \cite{Hirokawa1998,Brendza2000,Vale2003}.
Kinesin has two identical motor domains which are connected to each other by two strands called the neck linkers (NLs)(Figure 1). The NLs join together to form a coiled coil structure, which connects the motor to the cargo \cite{Vale2000}. Both motor domains can bind to the microtubule (MT). When both the heads are bound to a single protofilament, the motor domain that is closer to the plus end is referred to as the leading head (LH), and the one closer to the minus end is the trailing head (TH).

It is well established that kinesin walks towards the plus end by a hand-over-hand motion \cite{Asbury2003,Yildiz2004,Block07BJ}. According to this model, the  trailing head detaches from the MT and leaps over the leading head until it reattaches $\approx$16 nm further along the same protofilament \cite{Ray93JCB}, resulting in an effective step size of $\approx$8.2 nm. One ATP molecule is hydrolyzed per step. It is believed that a crucial initial event in kinesin motility involves a conformational change of the NL in the LH during which it docks to the LH \cite{Tomishige2006,Khalil2008}. The docking, which is a disorder to order transition in the NL of the LH, propels the TH predominantly  forward towards the target binding site on the plus end of the MT (in the absence of a resistive force).

Recent computational studies have shown that the conformational changes in the NL alone are not sufficient to enable the TH to reach the target binding site \cite{Hyeon2006,Hyeon2007,Zhang2012}. Our  previous study, which established that a single kinesin step can be broken into three distinct stages, showed that a significant amount of the stepping motion is due to tethered diffusion of the TH \cite{Zhang2012}. In order to dissect the kinematics of a single step of kinesin, we created a Coarse-Grained (CG) model of the complex of kinesin with MT \cite{Zhang2012}, and performed simulations of kinesin using Brownian dynamics with hydrodynamic interactions  \cite{Ermak1978} using the Rotne-Prager-Yamakawa diffusion tensor \cite{Rotne1969,Yamakawa1970}. We refer to this set of simulations as HI+. The inclusion of hydrodynamic interactions in simulations has been shown to dramatically improve the prediction of translational and rotational diffusion coefficients of proteins, and collapse kinetics \cite{Frembgen2009,Ando2010,Li12JPCB}. However, the calculations involving HI significantly increase the amount of time required to simulate large systems, such as the ones considered here, even if CG models are used.

Here, we performed CG simulations of kinesin without hydrodynamic interactions, (HI- simulations) to determine whether the inclusion of hydrodynamic interactions is necessary for a correct description of the kinesin stepping process. We demonstrate that it is important to include hydrodynamic interactions in order to simulate the complete stepping process within the expected experimental time frame for completing the step. In addition, we also show that changes in the diffusion coefficient of the motor domain have non-trivial effects on the stepping mechanism of kinesin, leading us to conclude that, although computationally intensive, it is crucial to include HI to quantitatively predict the stepping mechanism of motors in general, and kinesin in particular.

\section{Methods}
{\bf Model:} In order to simulate a large system consisting of the two motor domains and the polar track (MT), and the coiled coil it is necessary to use coarse-grained (CG) models. The use of CG models has been efficacious in making quantitative predictions for a number of problems in biology including the study of motors \cite{Hyeon2007,Hyeon11NatComm,Tehver2010,Whitford2010}. We used the Self-Organized-Polymer (SOP) Hamiltonian and simulated the stepping process using Brownian dynamics \cite{Zhang2012}. In the version of the SOP model used here, each amino acid is represented as a single bead centered around the $C_{\alpha}$ carbon \cite{Hyeon2006}. 

The SOP energy function is given by:
\begin{equation}
\begin{split}
H\left(r_{i}|X\right) & =V_{FENE}+V_{NB}^{A}+V_{NB}^{R}+V_{NB}^{E} \\
& = -\sum_{Y}\sum_{i=1}^{N^{Y}-1}\dfrac{k}{2}R_{0}^{2}\log\left(1-\dfrac{\left(r_{i,i+1}-r_{i,i+1}^{0}\left(X\right)\right)^{2}}{R_{0}^{2}}\right) \\
& +\sum_{Y,Y'}\sum_{i,j}\epsilon_{h}^{Y-Y'}\left[\left(\dfrac{r_{ij}^{0}\left(X\right)}{r_{ij}}\right)^{12}-2\left(\dfrac{r_{ij}^{0}\left(X\right)}{r_{ij}}\right)^{6}\right]\Delta_{ij} \\
& +\sum_{Y}\sum_{i=1}^{N^{Y}-2}\epsilon_{l}\left(\dfrac{\sigma}{r_{i,i+2}}\right)^{6}+\sum_{Y,Y'}\sum_{i,j}\epsilon_{l}\left(\dfrac{\sigma}{r_{ij}}\right)^{6}\left(1-\Delta_{ij}\right) \\
& +\sum_{Y,Y'}\sum_{i,j}\dfrac{q_{i}q_{j}}{4\pi \epsilon r_{ij}}e^{-\kappa r_{ij}}
\end{split}
\label{Model}
\end{equation}
where $r_{ij}$ is the distance between residues i and j. The first term in the Hamiltonian corresponds to the connectivity within each molecule and is given by the finite extensible non-linear elastic (FENE) potential. The second term accounts for attractive interactions between residues  in contact within the native structure. The third term is the repulsion potential for the remaining residues. Finally, the last term represents electrostatic interactions between charged residues. The labels Y and Y' refer to the residues in the molecules TH,LH, and MT and X refers to the state of the system. 

The two important energy parameters, $\epsilon_{h}^{LH-NL}$ and $\epsilon_{h}^{TH-MT}$, which determine the strength of the docking interactions of the NL associated with the LH and the TH-MT binding affinity, respectively play a key role in determining the motility of kinesin. Mutations in NL, which alter $\epsilon_{h}^{LH-NL}$, are known to impede the kinetics of stepping \cite{Khalil08PNAS}. The details of the model and the values of the different parameters can be found in the supplementary information of \cite{Zhang2012}. 

{\bf Dynamics:} The equations of motion are integrated using:
\begin{equation}
\textbf{r}_{i}(t+h)=\textbf{r}_{i}(t)+\dfrac{h}{\zeta}\textbf{F}_{i}+\boldsymbol\Gamma_{i}(t) ,
\end{equation}
where $r_{i}$ is the position of the $i^{th}$ residue represented using a single bead centered at the $C_{\alpha}$ position, $F_i$ is the force acting on residue i, $\zeta$ is the friction coefficient given by $\zeta = 6\pi\eta a$, and $\Gamma$ is a Gaussian random force which obeys,
\begin{equation}
\langle\boldsymbol\Gamma_{i}(t)\boldsymbol\Gamma_{j}(t')\rangle=6\dfrac{K_BT}{\zeta}h\delta_{ij}\delta_{tt'},
\end{equation}
For the simulations with hydrodynamic interactions, the corresponding equations of motion are,
\begin{equation}
\textbf{r}_{i}(t+h)=\textbf{r}_{i}(t)+\sum_j^N\dfrac{h\textbf{D}_{ij}}{K_BT}\cdot\textbf{F}_{j}+\boldsymbol\Gamma_{i}(t),
\end{equation}
In this case the random force obeys:
\begin{equation}
\langle\boldsymbol\Gamma_{i}(t)\boldsymbol\Gamma_{j}(t')\rangle=6\textbf{D}_{ij}h\delta_{tt'},
\end{equation}
where $\textbf{D}_{ij}$ is the Rotne-Prager-Yamakawa [9] [10] form of the diffusion tensor:
\begin{equation}
\textbf{D}_{ii}=\dfrac{(K_BT)}{6\pi\eta a}\textbf{I} ,
\end{equation}
\begin{equation}
\textbf{D}_{ij}=\dfrac{K_BT}{8\pi\eta r_{ij}}[(1+\dfrac{2a^2}{3r_{ij}^2})\textbf{I}+(1-\dfrac{2a^2}{r_{ij}^2})\dfrac{\textbf{r}_{ij}^T \textbf{r}_{ij}}{r_{ij}^2}]\qquad r_{ij}\geq 2a ,
\end{equation}
\begin{equation}
\textbf{D}_{ij}=\dfrac{K_BT}{8\pi\eta r_{ij}}[(\dfrac{r_{ij}}{2a})(\dfrac{8}{3}-\dfrac{3r_{ij}}{4a})\textbf{I}+\dfrac{r_{ij}}{4a}\dfrac{\textbf{r}_{ij}^T \textbf{r}_{ij}}{r_{ij}^2}]\qquad r_{ij}<2a .
\end{equation}

\textbf{Analyses:} In order to monitor the process of the NL docking to the LH we calculated
\begin{equation}
\Delta_{NL-LH}\left(t\right)=\sqrt{\dfrac{\sum_{\left(i,j\right)}\left(r_{ij}\left(t\right)-r_{ij}^{0}\right)^2}{N}} ,
\label{Delta}
\end{equation}
where $r_{ij}\left(t\right)$ is the distance between residue i in the NL and residue j in the docking site at time t, $r_{ij}^0$ is the distance between residues i and j in the crystal structure of the docked state (PDB code 2KIN), and N is the number of residue pairs involved in the docking interactions.

We followed the dynamics of the NL of the TH by calculating its end-to-end extension
\begin{equation}
X_{NL-TH}=\left|\bf{r}_{338}-\bf{r}_{326}\right| ,
\label{NL}
\end{equation}
where $\bf{r}_{338}$ and $\bf{r}_{326}$ are the positions of the TH residues at the ends of the NL.

\section{Results}
{\bf Diffusion is dominant in the initial stages in HI+ simulations:}
In order to quantify the effects of hydrodynamic interactions on the stepping kinetics of kinesin, we generated 100 HI+ trajectories for 6 $\mu s$ and 100 HI- trajectories for 18 $\mu s$. We set the docking strength parameter, $\epsilon_{LH-NL}$ (Eq. \ref{Model}), to 2 KCal/mol in order to ensure a stable docked state, which in turn enabled us to distinguish between the docking and the diffusive contributions to the motion of the TH. Experiment shows that although kinesin waits a long time between steps the duration of the step is only $\approx 20\mu s$ \cite{Carter2005}. Clearly, this is an average time for steps with a distribution that is Poissonian. In our HI+ simulations, we found that within 6 $\mu s$ up to 14$\%$ of the motors reached the target binding site, which is in good agreement with estimates based on the distribution of lifetimes \cite{Carter2005}. In sharp contrast, none of the HI- trajectories reached the target binding site within a 18 $\mu s$ window, indicating that indeed the inclusion of HI is necessary for correctly describing the motility of kinesin.

We computed the ensemble average displacement of the TH center of mass along the microtubule's axis (defined as the x axis), $\langle \Delta X_{TH}\rangle$, as a function of time, for each simulation type.
It can be seen clearly from our results (Figure 2a) that the TH in the HI+ simulations advances significantly faster compared to the HI- simulations. In both the HI+ and HI- simulations, the TH motion occurs in two stages, a rapid increase in $\langle \Delta X_{TH}\rangle$ followed by a slow increase. The most striking difference in the results between HI+ and HI- simulations is found in the fast phase. The time scale of initial increase in $\langle \Delta X_{TH}\rangle$ is shorter in the simulations with HI+ than in HI-. More importantly, there is substantial difference in the magnitude of displacement along the microtubule. In the HI- ensemble the TH moves on an average 5-6 nm during the fast phase, which accords well with the previous findings that the NL docking causes a 6 nm displacement of the TH. On the other hand, the TH mean displacement in the HI+ simulations is on the order of 10 nm (Figure 2a), about 4 nm further towards the plus end of the microtubule. It should be noted that even with HI the TH does not cover the required 16 nm. A third stage, discussed elsewhere \cite{Zhang2012}, involving interaction of the TH with the MT is needed for completing a single step. In \cite{Zhang2012} we established the critical role of MT in facilitating the completion of the 16 nm step. 

We dissected further the events in the fast phase by plotting a superposition of $\Delta X_{TH}$ in the individual trajectories (Figures 2b and 2c). The plots show that in the HI+ ensemble there are two dominant populations, one in the 4-7 nm range (shown in red in Figure 2(b)) and the other in the 12-16 nm range (blue in  Figure 2(b)). In the HI- ensemble only the 4-7 nm state is initially populated and the trajectories show slow transitions towards the second stage (Figure 2(c)). This is in contrast to the HI+ simulations where both stages occur during the fast phase, which explains the difference in the magnitudes of net displacement along the MT axis.

{\bf Neck linker dynamics in the absence of HI:}
In order to assess the effects of inclusion/exclusion of hydrodynamic interactions on the linker dynamics we measured both $\Delta_{NL-LH}$ (Eq. \ref{Delta}) and the TH linker's end-to-end distance, $X_{NL-TH}$ (Eq. \ref{NL}). The ensemble averages of these quantities as a function of time for each simulation type are in Figure 3. While the LH linker docking dynamics are almost identical in both cases (Figure 3 inset), there are significant differences between the two simulation types in $X_{NL-TH}$ (Figure 3 main panel). In both simulation types, the NL extension increases due to the rapid docking. In the HI+ simulations the increase is followed by a simple relaxation. In the HI- simulations relaxation is preceded by fluctuations that occur on $\approx0.4 \mu s$.

The difference in the NL dynamics can be explained by examining  the pathways taken by the TH in each case. We calculated the ensemble average position of the TH center of mass in the xz plane, $\langle \textbf{r}_{xz} \rangle$, for both simulation types, at different times (Figure 4a). When comparing the two trajectories, it can be seen that the HI- trajectory is initially closer to the microtubule in comparison to the HI+ trajectory (compare green segment and the gray path in Figure 4a). The difference arises because the TH motion in the HI- simulations is dominated by the NL docking. In the HI+ case, on the other hand, the effects of diffusion of the trailing head are significant, which push the TH further away from the microtubule. The end result is that in the HI- simulations, the TH comes into contact with the bound LH, which likely causes internal friction slowing down the relaxation of the neck linker (Figures 4a and 4b). The distance between the centers of mass of the two motor heads as a function of $t$  vividly illustrates (Figure 5) that in the presence of HI+ interactions  the $d_{TH-LH}$ is considerably greater that in HI- simulations. A corollary of this finding is that mutations in the NL of the motor heads that would over stabilize the strength of the LH-NL interaction should slow down mobility even in the presence of HI.

\section{Discussion}
In this study we demonstrated the significant role that hydrodynamic interactions play in the kinesin stepping process. In particular, our results firmly establish that in order to obtain realistic diffusion times for the TH stepping, hydrodynamic interactions have to be included in the calculations. While in HI+ simulations the TH reaches the target binding site within realistic time scales, it fails to do so in the HI- case. Because we did not observe a single binding event within the observed experimental time scale for completing the step we surmise that HI- simulations are not useful in describing motor dynamics. We note parenthetically that HI interactions are not relevant when considering equilibrium properties such as the distribution of the center of mass of the TH.

Further comparison between simulations with and without hydrodynamic calculations allowed us to explore the effects of the diffusive properties of the TH on the stepping dynamics. The striking result is that slowing down the TH diffusion results in non trivial dynamics when NL docking is rapid. This is potential due to the frictional effects of the LH on the TH when the two come into contact to produce the docked state. This results in rugged behavior of the  end-to-end extension of the NL, and a slower initial mobility. The inclusion of HI speeds up diffusion resulting in much higher probability that the trailing head reaches the target binding sites within several microseconds, in accord with experiments. Although demonstrated within the context of stepping kinetics of kinesin, we believe that the conclusions must also hold for motility of other (myosin and dynein) motors as well. 

{\bf Acknowledgements:} This work was supported by a grant from the National Science Foundation through grant number CHE 13-61946.

\newpage
{\bf FIGURE CAPTIONS}

%\bibliography{refer1}
%\bibliographystyle{unsrt}

\begin{figure}[h]
	\centering
 \includegraphics[width=0.9\textwidth]{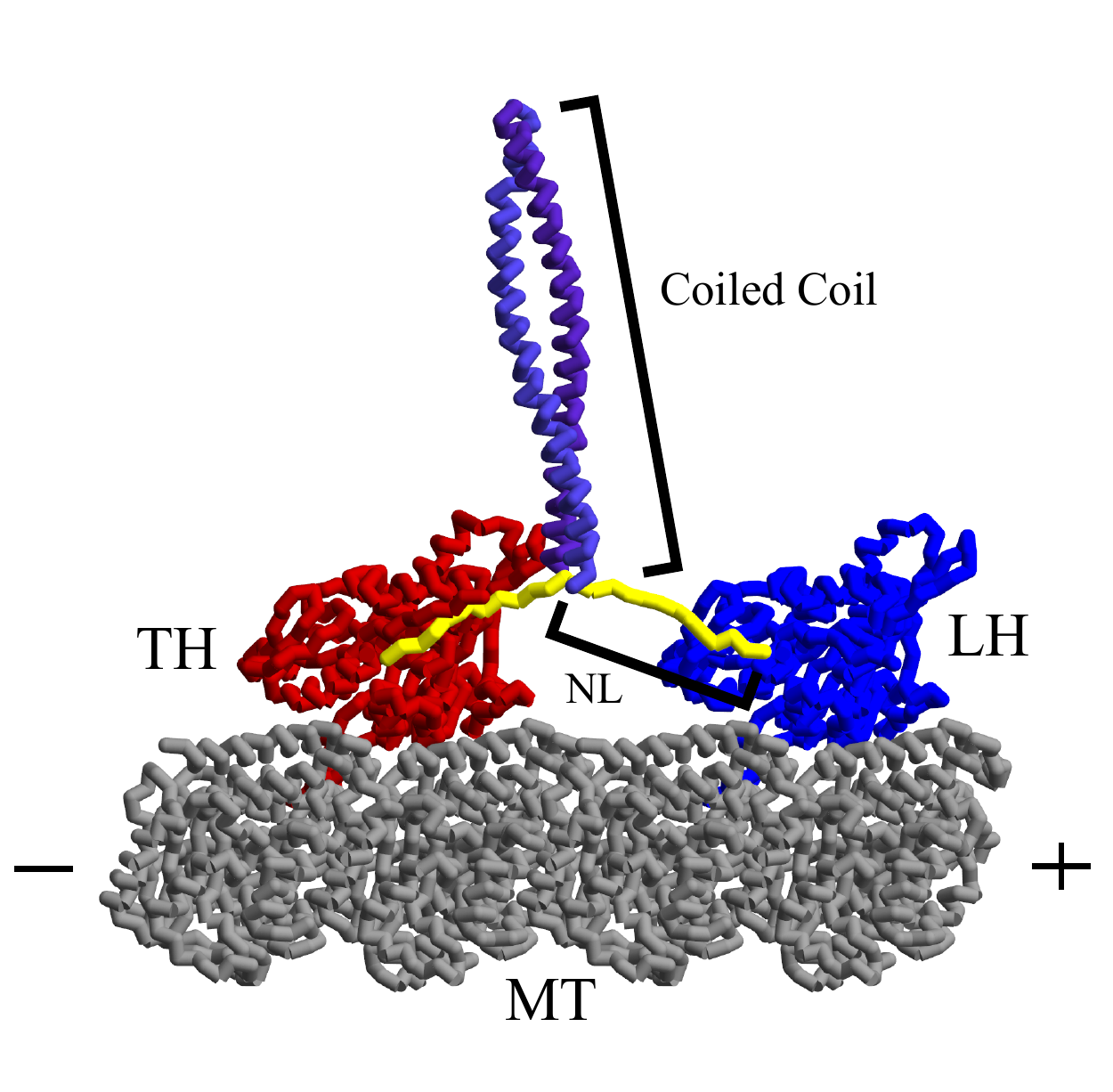}
	\caption{Structure of the motor and MT complex used in the simulations. Components by color: MT-gray, TH-red, LH-blue, NL-yellow, coiled coil-purple. The structure of the complex was constructed using existing data for some of the components. The procedure used to obtain the complex is given in \cite{Zhang2012}.}
\end{figure}

\begin{figure*}[h]
\centering
%\begin{subfigure}[t]{0.5\textwidth}
\includegraphics[width=0.5\textwidth]{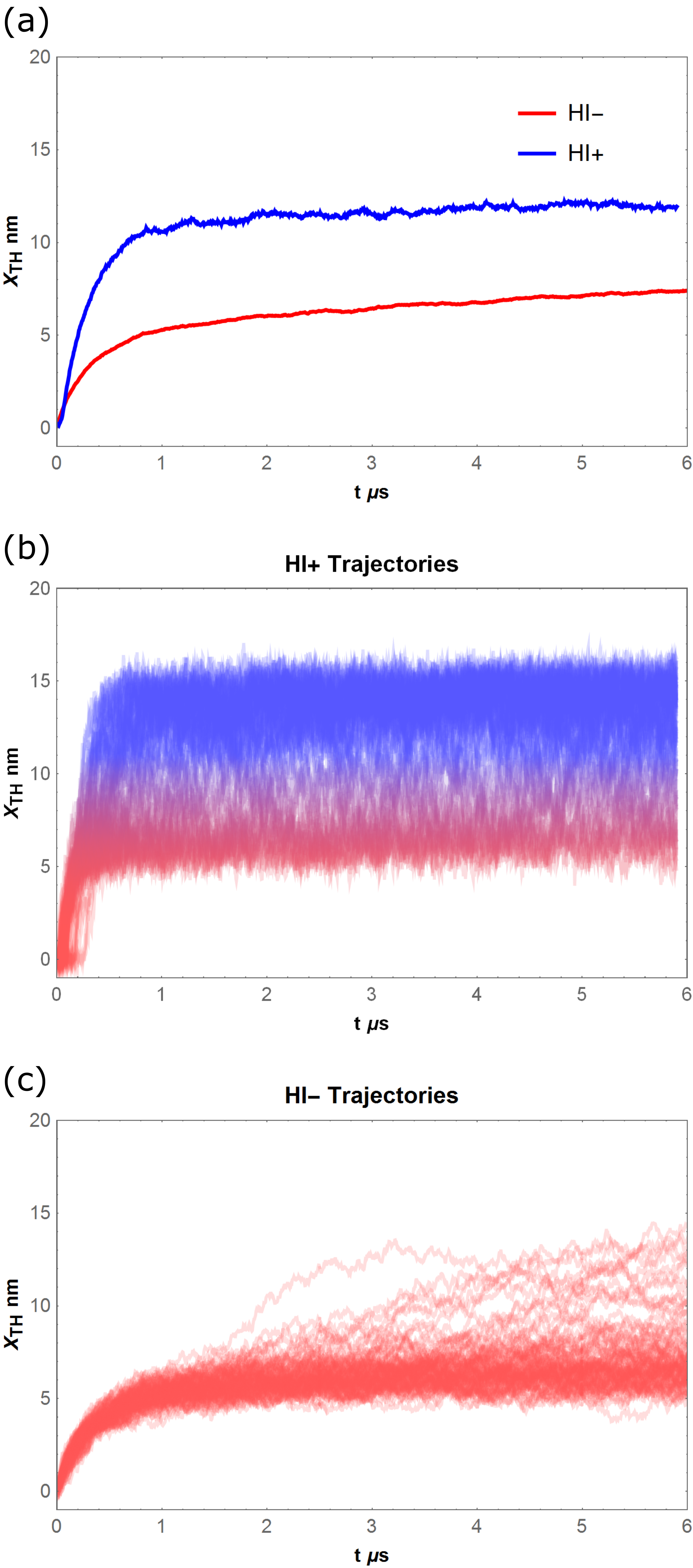}
%\centering
%\begin{subfigure}[t]{0.5\textwidth}
%\includegraphics[width=\textwidth]{Fig2New.pdf}
%\caption{}
%\end{subfigure}
%\begin{subfigure}[t]{0.5\textwidth}
%	\includegraphics[width=\textwidth]{Fig2bAlt.png}
%	\caption{}
%\end{subfigure}
%\begin{subfigure}[t]{0.5\textwidth}
%	\includegraphics[width=\textwidth]{Fig2c.png}
%	\caption{}
%\end{subfigure}
\caption{
(a) Mean displacement of the center of mass of the TH from its initial position projected along the MT axis as a function of time. The red curve corresponds to the HI+ simulations. (b) Superposition of the different HI+ trajectories showing $X_{TH}\left(t\right)$ as a function of time. (c) Same as (b) except the trajectories are from simulations without HI.}
\end{figure*}

\begin{figure*}[h]
	\centering
	\includegraphics[width=\textwidth]{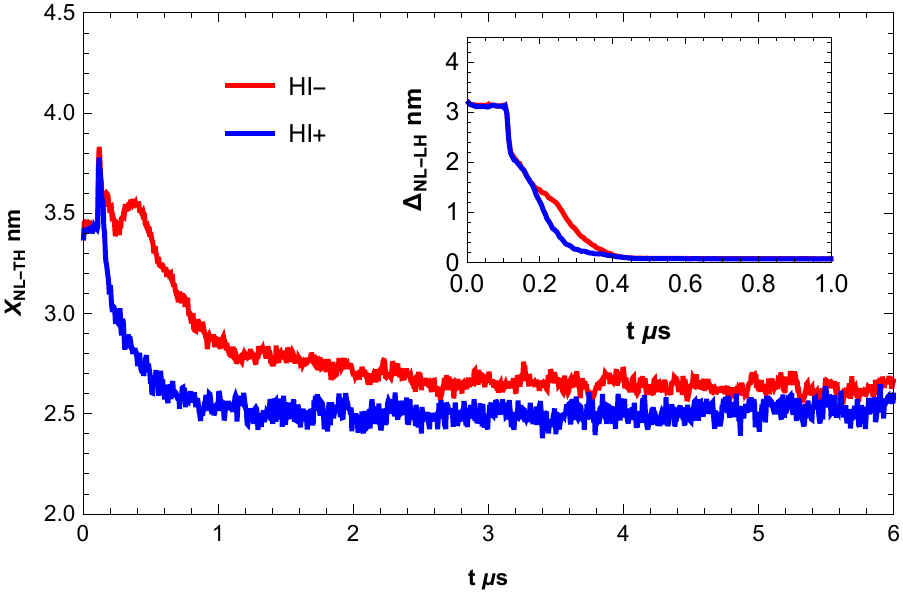}
	\caption{
	Mean values of the end-to-end extension of the NL of the TH, $X_{NL-TH}$ (Eq. \ref{NL}), as a function of time for both HI+ and HI- ensembles. Inset shows the average values of $\Delta_{NL-LH}$ (Eq. \ref{Delta}), a probe of the NL docking process, as a function of time for both HI+ and HI- ensembles.}
\end{figure*}

\begin{figure*}[h]
\centering
	\includegraphics[width=0.7\textwidth]{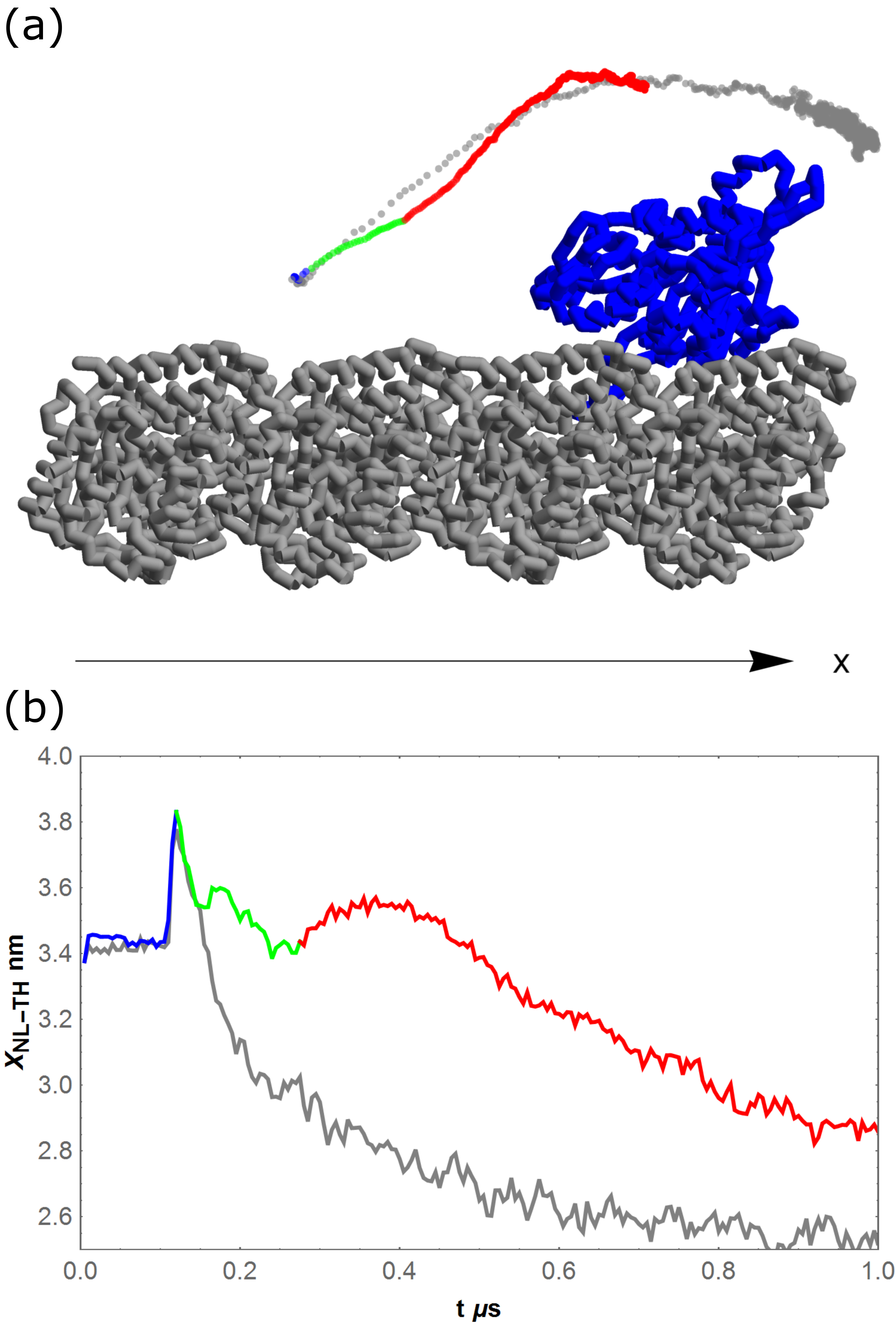}
    \caption{
	(a) The mean trajectory of the TH center of mass, projected onto the xz plane. The MT (gray) is along the x axis and the LH (blue) is tightly bound to the MT. The colored path corresponds to the HI- ensemble and the gray path corresponds to the HI+ ensemble. The different colors in the HI- path highlight the corresponding time domains in (b). (b) Mean values of $X_{NL-TH}$ as a function of time. The colored curve corresponds to the HI- ensemble and the gray curve corresponds to the HI+ ensemble.}
\end{figure*}

\begin{figure}[h]
	\centering
	\includegraphics[width=\textwidth]{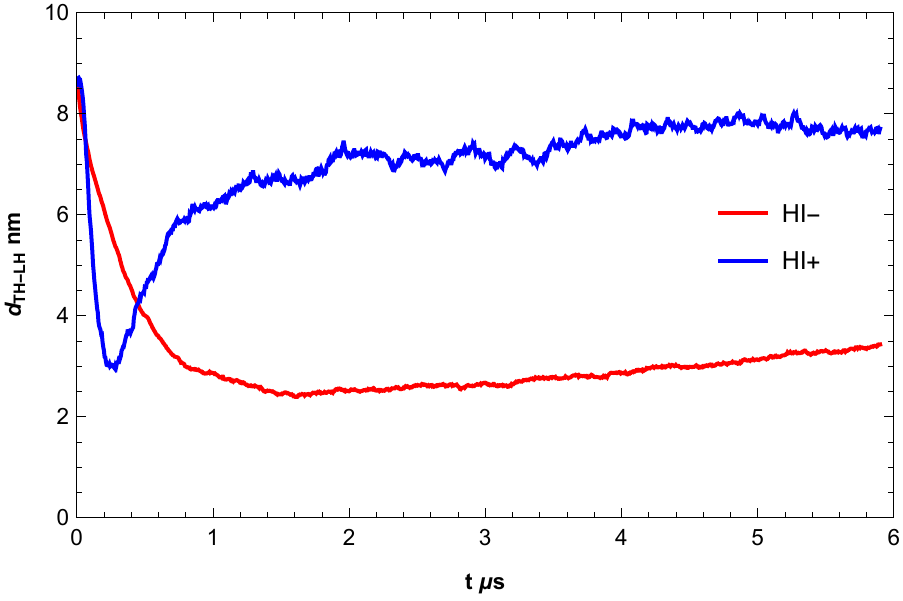}
	\caption{Mean distance between the TH and LH centers of mass as a function of time. The blue curve corresponds to the HI+ simulations and the red curve corresponds to the HI- simulations. Inclusion of hydrodynamic interactions qualitatively change the results.}
\end{figure}

\end{document}